%% file: convergencePRA.tex
\documentclass[%
preprint,
amsmath,amssymb,
aps,
pra,
showkeys
]{revtex4-2}

\usepackage{graphicx}
\usepackage{dcolumn}
\usepackage{bm}

\begin{document}
\title{Existence Criterion of Solutions to the Inverse Problem of Photocount
Statistics Obtained by the Inverse Bernoulli Transform}
\author{Pavel P. Gostev}
\email{gostev.pavel@physics.msu.ru}

\author{Sergey A. Magnitskiy}
\email{sergeymagnitskiy@gmail.com}

\author{Anatoly S. Chirkin}
\email{aschirkin@physics.msu.ru}

\affiliation{Lomonosov Moscow State University}

\begin{abstract}
It is shown that the applicability conditions for the inverse Bernoulli
transform method for solving the inverse problem of photocount statistics
are determined by the fulfillment of the associativity condition for
multiplying the matrices included in this transformation. A general
criterion for evaluating the photocount distributions $Q_{m}$ in
the case of few-photon light, which makes it possible to establish
whether the solution to the inverse problem of photocount statistics
by inverse Bernoulli transform method is applicable for $\eta<0.5$,
is found. As an example of application of the obtained criterion,
the critical quantum efficiency $\eta_{cr}$ is found for compound
Poisson distribution, below which the solution of the inverse problem
of photocount statistics becomes incorrect. Additionally it is shown
that the normalization of $Q_{m}$ is not sufficient to obtain a correct
solution using the inverse Bernoulli transform. 
\end{abstract}

\maketitle

\section{Introduction}

One of the most widely used methods for determining the energy characteristics
of light is based on measuring the photocount distribution~\cite{britun2018photon},
that is, the statistics of electrons emitted from the photocathode
irradiated by the light beam. This method is based on detecting the
optical radiation by single-photon counters~\cite{eisaman2011invited}.
To date, the photon counting method, which has a long history~\cite{morel2005100},
is widely used in both applied~\cite{wick1989photon,hadfield2009single,taguchi2013vision,chunnilall2014metrology,flohr2020photon}
and fundamental research~\cite{banaszek1996direct,banaszek1999direct,nehra2019state,ortolano2021experimental}.
Nowadays, it is one of the key experimental methods used in quantum
optics.

\subsection{Semi-classical inverse problem of photocount statistics}

Already at the initial stage of application of the photon counting
method, in addition to the direct problem --- finding the photocount
distribution from the known light state, interest was aroused by the
inverse problem --- determination of the light properties from the
known photocount statistics. One of the most important properties
of light is the energy distribution. For the first time, the problem
of reconstructing the energy distribution from the photocount one
was considered in~\cite{wolf1964determination}; later, various approaches
were developed to solve this problem~\cite{bedard1967light,sultani1995inverse,earnshaw1996inversion}.
These studies were based on Mandel's semi-classical formula for the
photocount distribution $Q_{m}$~\cite{mandel1958fluctuations},
which is mathematically the averaged Poisson distribution over the
energy distribution of the detected radiation (Poisson transformation
of the energy distribution):

\begin{equation}
Q_{m}=\int_{0}^{\infty}\frac{(\eta\mathcal{E})^{m}}{m!}\exp(-\eta\mathcal{E})w(\mathcal{E})d\mathcal{E},\label{eq:mandel}
\end{equation}
where $\mathcal{E}=\mathcal{E}\left(T\right)=\int_{t}^{t+T}\int_{S}I(\vec{r},t)d^{2}rdt$
is the light energy falling onto the detector area $S$ during time
$T$; $\eta$ is the quantum efficiency of detection; $I(\vec{r},t)$
is the light intensity; $w(\mathcal{E})$ is the probability density
of the fluctuating parameter $\mathcal{E}$; $m$ is a number of photoelectrons,
emitted during time $T$.

Within the framework of the semi-classical model, the inverse problem
of photocount statistics~\cite{akhmanov1981introduction} is the
reconstruction of the distribution $w(\mathcal{E})$ from measured
distribution $Q_{m}$. The problem of reconstructing the energy distribution
has been solved more than once. Thus, in~\cite{wolf1964determination},
the Poisson transform was inverted using the Fourier transform and
the apparatus of characteristic functions. In~\cite{bedard1967light},
an expansion of the intensity distribution in terms of Laguerre polynomials
was applied. The authors of~\cite{sultani1995inverse} used Pad{\'e}
approximants to inverse the Poisson transform, and
cubic B-splines are used in~\cite{earnshaw1996inversion} for the same purpose.

\subsection{Quantum inverse problem of photocount statistics}

For a correct description of the photodetection process, especially
when applied to few-photon light, the field itself should also be
considered as a quantum object. A consistent theory of such a process
gives the following result~\cite{vogel2006quantum}:

\begin{equation}
Q_{m}=\left\langle :\dfrac{1}{m!}\left[\eta\hat{\mathcal{E}}\right]^{m}\exp\left[-\eta\hat{\mathcal{E}}\right]:\right\rangle .\label{eq:7-1}
\end{equation}
Here $\hat{\mathcal{E}}=\hat{\mathcal{E}}(T)=\int_{t}^{t+T}\int_{S}\hat{E}^{-}(\boldsymbol{r},t)\hat{E}^{+}(\boldsymbol{r},t)d^{2}rdt$;
$\hat{E}^{-}(\hat{E}^{+})$ is the negative- (positive-) frequency
field operator, $\left\langle ::\right\rangle $ is the normally ordered
averaging. As shown in~\cite{vogel2006quantum}, the expression~(\ref{eq:7-1})
can be rewritten in the Fock basis as

\begin{equation}
Q_{m}=\sum_{n=m}^{\infty}C_{n}^{m}\eta^{m}(1-\eta)^{n-m}P_{n},\label{eq:btransform}
\end{equation}
where $P_{n}=\left\langle :\dfrac{1}{m!}\hat{\mathcal{E}}^{m}\exp\left[-\hat{\mathcal{E}}\right]:\right\rangle $
is the photon-number distribution, or the probability that $n$ photons
hits a detector during a time $T$, and $C_{n}^{m}=n!/[m!(n-m)!]$
is the binomial coefficient. The formula~(\ref{eq:btransform}) is
commonly called the Bernoulli transformation~\cite{lee1993external,kiss1995compensation,herzog1996loss}.
The distribution~(\ref{eq:btransform}) allows us to give a clear
physical interpretation. Since $\eta$ is the probability of registering
one photon during the measurement interval, then $\eta^{m}$ is the
probability of registering $m$ photons, and $(1-\eta)^{(n-m)}$ is
the probability of not registering $(n-m)$ photons when $n$ photons
arrive at the detector. The coefficient $C_{n}^{m}$ takes into account
the possible number of combinations of occurrence of $m$ photoelectrons.

Thus, in the quantum approach, the photocount distribution is given
by the convolution of the photon-number distribution with the Bernoulli
one~(\ref{eq:btransform}) and differs from the semi-classical Mandel
formula~(\ref{eq:mandel}). Therefore, at a low-intensity (few-photon) level, for a correct description of the photodetection
process, one must proceed from the Bernoulli transformation~(\ref{eq:btransform}).

The inverse problem of the photocount statistics in a few-photon mode
is the finding $P_{n}$ from the known distribution $Q_{m}$ which
are associated by the relation~(\ref{eq:btransform}). The importance
of this problem lies, in particular, in the fact that currently, few-photon
light sources are of significant interest in quantum technologies~\cite{messin2009few,eisaman2011invited,pathak2018classical}.
Photon-number distribution for few-photon radiation is an analog of
the intensity for bright radiation; therefore, it can be considered
as one of the most important characteristics of a light source. Experimentally,
the photon counting method is apparently the simplest and cheapest
method to get data about the photon-number distribution. If we had
an ideal photon counter with $\eta=1$ there wouldn't be any problem
with obtaining these data, as distributions $Q_{m}$ and $P_{n}$
coincide. However, the quantum efficiency of existing photon counters
is typically less than 1, for example, for modern SiPM detectors $\eta\le0.65$~\cite{gundacker2020silicon}.
This, as will be shown below, causes significant difficulties in inverting
the formula~(\ref{eq:btransform}).

At present, numerical statistical methods are most often used to solve
the inverse problem. They give an approximate solution to the problem
under some reasonable assumptions about its structure. One of the
main methods of this kind is the maximum likelihood method~\cite{aitchison1958maximum},
in which a solution is sought that maximizes the likelihood function.
Combined with expectation~\cite{zambra2006reconstruction,zambra2007nontrivial}
and entropy~\cite{hlouvsek2019accurate} maximization methods, it
gives very good results. However, these methods are not universal.
The problem is that they are approximate and use \textit{a priori}
information when searching for a solution, which limits their generality,
making it impossible to solve the problem reliably in all cases.

Analytical methods~\cite{lee1993external,kiss1995compensation,herzog1996loss,wunsche1990reconstruction,herzog1996generating}
are free from this shortcoming. Analytical methods, in contrast to
numerical ones, not only give an exact solution to the problem, but
also allow us to understand its essence. The best known and most commonly
used analytical method for solving the inverse problem for few-photon
light, which takes into account the losses in photodetection due to
$\eta<1$, is the method based on the direct inversion of the Bernoulli
transformation~\cite{lee1993external,kiss1995compensation}. As is
known, one of the main fundamental problems that arise in solving
inverse problems is the problem of stability of their solutions. Apparently,
the first work in which a fundamental study of the stability of solutions
obtained by the inverse Bernouli transform was carried out is the
work~\cite{kiss1995compensation}, where the authors managed to get
an expression of statistical uncertainty for $P_{n}$ values and to
make the first base research of the solution convergence. They showed
that for any finite $Q_{m}$ the solutions are stable for arbitrary
$\eta$. However, for infinite $Q_{m}$, the solutions are stable
only if $\eta>0.5$, and in the case of $\eta<0.5$, one can find
counterexamples when the solution turns out to be unstable. As an
example, they gave the thermal distribution, for which the stable
reconstruction is impossible below some critical quantum efficiency.
The results obtained in this work caused a scientific discussion~\cite{d1998loss,kiss1998reply,d1997comment,kiss1998replyoncomm}
about whether the threshold value identified in~\cite{kiss1995compensation}
is a fundamental limitation, i.e. whether it is due to any physical
reasons or due to mathematical problems of the solution. However,
the causes for the instability of solutions were not identified in
the subsequent works, and the question remained open.

In this paper, we managed to find the reasons for the apparent instability
of the solutions of the inverse Bernoulli transform. In addition,
we have found a general criterion for evaluating the distributions
$Q_{m}$, which allows us to establish whether the solution of the
inverse problem obtained by the inverse Bernoulli transform method
is mathematically correct for a given value of $\eta$ less then 0.5.

\section{Inverse Bernoulli transform method }

The inverse Bernoulli transform method is based on the fact that the
formula~(\ref{eq:btransform}) can be reversed~\cite{lee1993external}.
The easiest way to see this is to represent this formula in matrix
form $\mathbf{Q}=\hat{T}\mathbf{P}$. In the case of a finite $P_{n}$,
when the number of members in the distribution is limited to $N$,
the matrix representation of the formula~(\ref{eq:btransform}) has
the form:
\begin{widetext}
\begin{equation}
\left(\begin{array}{c}
Q_{0}\\
Q_{1}\\
.\\
.\\
Q_{N-1}\\
Q_{N}
\end{array}\right)=\left(\begin{array}{ccccccc}
1\quad & (1-\eta)\quad & (1-\eta)^{2} & \cdot & \cdot & (1-\eta)^{N}\\
0 & \eta & 2\eta(1-\eta) & \cdot & \cdot & N\eta(1-\eta)^{N-1}\\
0 & 0 & \eta^{2} & \cdot & \cdot & \cdot\\
\cdot & \cdot & \cdot & \ddots & \cdot & \cdot\\
0 & 0 & 0 & . & \eta^{N-1} & N\eta^{N}(1-\eta)\\
0 & 0 & 0 & \cdot & 0 & \eta^{N}
\end{array}\right)\left(\begin{array}{c}
P_{0}\\
P_{1}\\
\cdot\\
\cdot\\
P_{N-1}\\
P_{N}
\end{array}\right),\label{eq:1-2}
\end{equation}
\end{widetext}

\noindent or simply 
\begin{equation}
\mathbf{Q}=\hat{T}\mathbf{P},\label{eq:mdirect}
\end{equation}
where $\mathbf{Q}$ and $\mathbf{P}$ are the N-dimensional vectors.
This matrix equation is easy to solve: 
\begin{equation}
\mathbf{P}=\hat{T}^{-1}\mathbf{Q},\label{eq:minverse}
\end{equation}
where $\hat{T}^{-1}$ is an inverse matrix to $\hat{T}$.

\noindent Calculating the inverse matrix and passing back to analytical
form, it is possible to show~\cite{lee1993external} that the solution
to the inverse problem of photocount statistics can be represented
as

\begin{equation}
P_{n}=\sum_{m=n}^{\infty}C_{m}^{n}\eta^{-n}(1-\frac{1}{\eta})^{m-n}Q_{m}.\label{eq:invb}
\end{equation}

As seen from the formula~(\ref{eq:invb}), the inverse matrix is
triangular, like the original matrix $\hat{T}$. If the distribution
$Q_{m}$ is finite and no restrictions are imposed on it, then no
problems with the solutions to the inverse problem arise for any $\eta$~\cite{kiss1995compensation}. 

\begin{figure}[b]
\centering\includegraphics[width=1\linewidth]{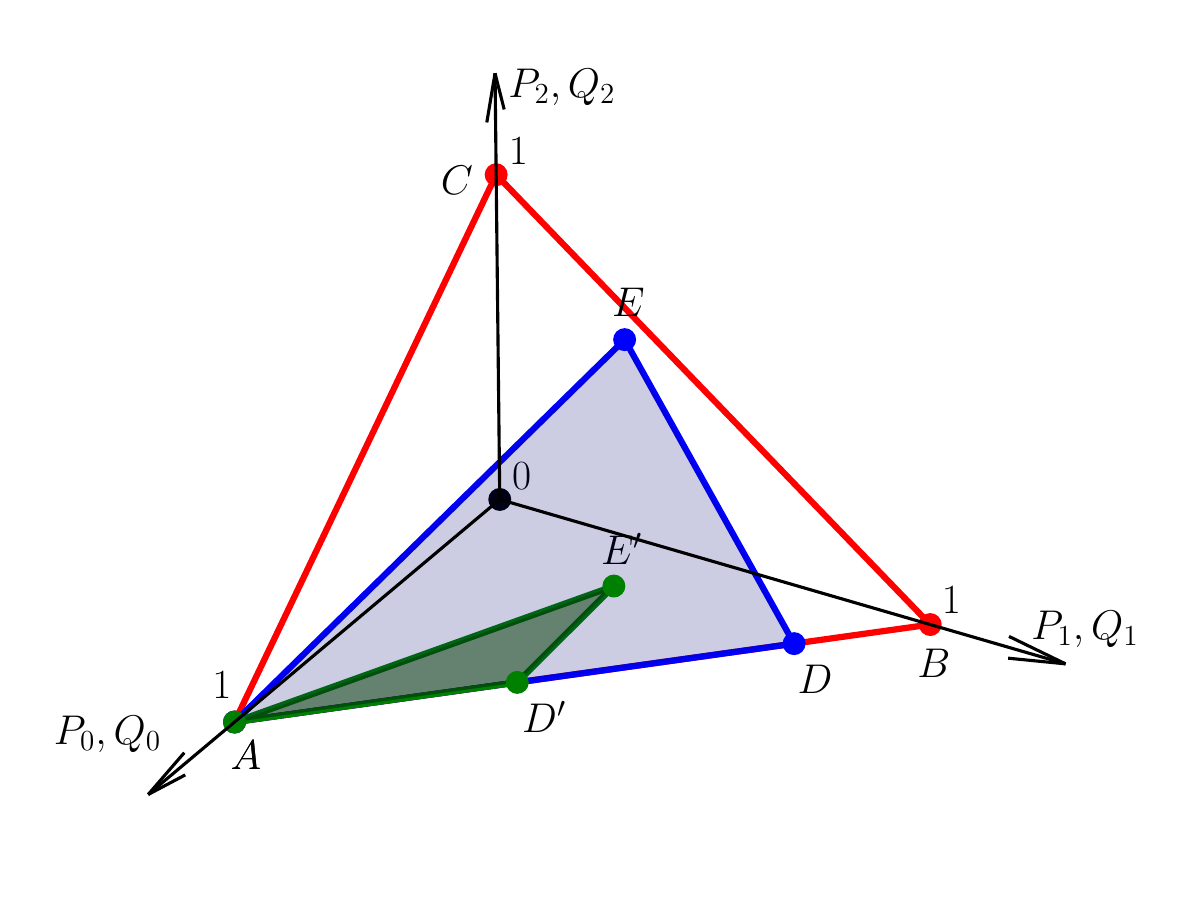}
\caption{\label{fig:1} Ranges of possible values of $P_{n}$ (red) and $Q_{m}$
for $\eta=0.8$ (blue), and $\eta=0.4$ (green), depicted in 3-dimensional
space.}
\end{figure}

Before turning to the presentation of the main results related to
the analysis of the inverse Bernoulli transform as applied to infinite
distributions of photons and electrons, let us dwell on some important
properties of this transform, which usually remain outside the scope
of the standard consideration. Firstly note that vectors $\mathbf{P}$
and $\mathbf{Q}$ are the probability distributions, hence $P_{n}\geq0$
for all $n$ and $Q_{m}\geq0$ for all $m$. Moreover, the normalization
conditions must be met, i.e. $\sum_{n=0}^{N}P_{n}=1$ and $\sum_{m=0}^{N}Q_{m}=1$.
The listed restrictions imposed on $P_{n}$ and $Q_{m}$ lead to the
important feature of the solution~(\ref{eq:invb}), namely that the
transformation $\hat{T}$ turns out to be contracting. If to interpret
$P_{n}$ and $Q_{m}$ as projections of unit vectors to coordinate
axes in $N$-dimensional space, it is convenient to illustrate a contraction
action of $\hat{T}$ by depicting the ranges of possible values for
$P_{n}$ and $Q_{m}$ in such space. To illustrate this, refer to
Fig.~\ref{fig:1}, which shows the ranges of valid values of $P_{m}$
and $Q_{m}$ for 3-dimensional distributions. In Fig.~\ref{fig:1}
the distributions $P_{n}$ and $Q_{m}$ are interpreted as projections
of unit vectors on the coordinate axes in 3-dimensional space.

The domains of possible values of $P_{n}$ and $Q_{m}$ lie in the
same plane, which intersects the coordinate axes at the points $A=\left\{ 1,0,0\right\} ,B=\left\{ 0,1,0\right\} ,C=\left\{ 0,0,1\right\} $.
However, the sizes of these domains are different. Thus, the possible
values of $P_{n}$ fill the equilateral triangle $ABC$, while the
possible values of $Q_{m}$ lie inside the triangle $ADE$ or $AD^{\prime}E^{\prime}$
of a smaller areas depending on $\eta$. Note that the coordinates
$D$ and $E$ of the vertices of this triangle coincide with the columns
of the matrix $\hat{T}$, and, therefore, depend on the $\eta$: $D=\left\{ 1-\eta,\eta,0\right\} ,E=\left\{ (1-\eta)^{2},2\eta(1-\eta),\eta^{2}\right\} $.
If $\eta=1$ then P- and Q-triangles coincide. In the case of $N$
dimensions the reasoning is similar, but instead of triangles one
should use multi-dimensional generalization of a triangle --- simplex.

This specified property of the transformation $\hat{T}$ causes potential
incorrectness in the $P_{n}$ reconstruction from experimentally obtained
$Q_{m}$ according to the formula~(\ref{eq:invb}). Indeed, in the
case of a contraction mapping, normalization of the $Q$ distribution
does not guarantee the correctness of reconstruction, since although
the normalization guarantees that the end of the vector $\mathbf{Q}$
falls on the $ABC$ plane, it does not guarantee that it falls inside
the Q-simplex. As a result, the reconstructed $P_{n}$ distribution
may be outside of P-simplex and you can get, in principle, arbitrary
values of $P_{n}$, both greater than 1 and even negative.

To illustrate this conclusion, consider an example of incorrect recovery
of $P_{n}$ from a pre-normalized 3-dimensional $Q_{m}$. Let us assume
that $\mathbf{Q}=\left\{ 0,0,1\right\} $, then according to formula~(\ref{eq:invb})
the elements of the reconstructed distribution $P_{n}$ will have
the form: $P_{0}=(1-\frac{1}{\eta})^{2}$, $P_{1}=2\eta^{-1}(1-\frac{1}{\eta})$,
$P_{2}=\eta^{-2}$. For example, if $\eta=0.5$ then $\mathbf{P}=\left\{ 1,-4,4\right\} $.
Note that despite the obtained negative value of $P_{1}$, the normalization
of the distribution $P_{n}$ remains.

This feature of the solution~(\ref{eq:invb}) becomes critical in
the processing of experimental data and, no doubt, requires the development
of special methods which would not allow the experimental values to
go beyond the boundaries of the Q-simplex. However, this problem is
beyond the scope of the questions discussed in this article, and we
will not touch on this issue further.

\section{Criterion for the existence of a solution to the inverse problem
of photocount statistics for infinite distributions}

In most physical problems, the photocount statistics is described
by infinite distributions. As mentioned in the introduction, when
studying the properties of the solution to the inverse problem of
photocounts by the method of inverse Bernoulli transformation in the
case of infinite distributions, a possible instability of the behavior
of the solution~(\ref{eq:invb}) was found for quantum efficiency
values $\eta<0.5$. At the same time, in the case of finite distributions,
no features in the behavior of solutions were noted. If we trace the
derivation of the formula~(\ref{eq:invb}), then at first glance,
no mathematical reasons for the appearance of the instability threshold
are visible. Indeed, let us describe in more detail the procedure
for finding a solution to the inverse problem in matrix form~(\ref{eq:mdirect}).
Multiplying both sides of the equation~(\ref{eq:mdirect}) on the
left by $\hat{T}^{-1}$, we get $\hat{T}^{-1}\mathbf{Q}=\hat{T}^{-1}\hat{T}\mathbf{P}$.
Considering that $\hat{T}^{-1}\hat{T}=1$, we directly arrive at the
solution of the inverse problem: $\mathbf{P}=\hat{T}^{-1}\mathbf{Q}$.
But, if we look closely at the above derivation, we can see that the
derivation implicitly assumed the associativity of matrix multiplication
$\hat{T}^{-1}$, $\hat{T}$ and $\mathbf{P}$. But, as follows from
functional analysis, the product of infinite matrices is generally
not associative, i.e. $\hat{T}^{-1}\left(\hat{T}\mathbf{P}\right)$
is not always equal to $\left(\hat{T}^{-1}\hat{T}\right)\mathbf{P}$.
It follows that the solution to the problem formulated in the introduction
is not to find the stability regions of the solution~(\ref{eq:minverse}),
but to determine the ranges of $\eta$ for which the matrix product
$\hat{T}^{-1}\hat{T}\mathbf{P}$ is associative, i.e. when the solution
of the inverse problem can in principle be written as~(\ref{eq:minverse}).

As an associativity criterion, one can choose the conditions for the
convergence of the series that determine the elements of the products
of matrices. From a technical point of view, it is easier to examine
not the associativity of matrices, but to examine for convergence
the final solution~(\ref{eq:invb}), which from a mathematical point
of view can be viewed as a countable set of series. Therefore, the
solution exists if all series~(\ref{eq:invb}) converge for any n.

Below are the results of studying the existence of a solution to the
inverse problem in the form~(\ref{eq:minverse}) for various distributions
of photons. Let's choose from the solution~(\ref{eq:invb}) an arbitrary
series with number $n$ and rewrite it in the equivalent form

\begin{equation}
P_{n}=(\eta-1)^{-n}\sum_{m=n}^{\infty}\left(-1\right)^{m}(\frac{1}{\eta}-1)^{m}C_{m}^{n}Q_{m},\label{eq:invb_equiv}
\end{equation}
whence it can be seen that it is an alternating series. Denoting $a_{nm}=(\frac{1}{\eta}-1)^{m}C_{m}^{n}Q_{m}$,
we can write it in a more compact form:

\begin{equation}
P_{n}=(\eta-1)^{-n}\sum_{m=n}^{\infty}\left(-1\right)^{m}a_{nm}.
\end{equation}

An interesting feature of the solution~(\ref{eq:invb_equiv}) is
the presence of a critical value of the quantum detection efficiency
$\eta_{cr}=0.5$. You can see this by looking at the structure of
the sequence $a_{nm}$, which can be considered as the product of
two sequences $a_{nm}^{\left(1\right)}=(\frac{1}{\eta}-1)^{m}C_{m}^{n}$
and $a_{m}^{\left(2\right)}=Q_{m}$. Because the series $\sum_{m=n}^{\infty}Q_{m}$
converges for any distributions $Q_{m}$ due to the normalization
condition, then by Abel convergence criterion it is sufficient for
the convergence of the series~(\ref{eq:invb_equiv}) that the sequence
$a_{nm}^{\left(1\right)}$ was monotone and limited. The sequence
$a_{nm}^{\left(1\right)}$ starting from some number $m$ is monotone,
therefore, for its boundedness it is sufficient that it converges
to $0$. As follows from the explicit form of $a_{nm}^{\left(1\right)}$,
for $\eta>0.5$ $\underset{m\rightarrow\infty}{\lim}a_{nm}^{\left(1\right)}=0$,
and for $\eta<0.5$ $\underset{m\rightarrow\infty}{\lim}a_{nm}^{\left(1\right)}=\infty$.
This implies that for $\eta>0.5$ the series (\ref{eq:invb_equiv})
converges for any distribution $Q_{m}$, and for $\eta<0.5$ the convergence
of the series (\ref{eq:invb_equiv}) depends on the type of distribution
$Q_{m}$.

Now we can find the criterion that the distribution $Q_{m}$ must
satisfy to ensure the convergence of the series~(\ref{eq:invb_equiv})
for $\eta<0.5$. The series~(\ref{eq:invb_equiv}) will converge
if the Leibniz criterion is fulfilled, i.e. the sequence $a_{nm}$
monotonically tends to zero as $m\rightarrow\infty$. Because the
sequence $a_{nm}$ is non-negative, it is sufficient to satisfy the
monotonicity condition $a_{n,m+1}<a_{nm}$ starting from some number
$M$. Noticing that 
\begin{equation}
a_{n,m+1}=a_{nm}\dfrac{m+1}{m-n+1}\left(\frac{1}{\eta}-1\right)\dfrac{Q_{m+1}}{Q_{m}},
\end{equation}
we obtain the condition of monotonicity for the sequence $a_{nm}$
in the form:

\begin{equation}
Q_{m+1}<Q_{m}\cdot\left(1-\frac{n}{m+1}\right)\frac{\eta}{1-\eta}.\label{eq:conv_criterion}
\end{equation}

The relation~(\ref{eq:conv_criterion}) can be regarded as a criterion
for the convergence of the series~(\ref{eq:invb}) for an arbitrary
$n$. In relation to the question of the existence of the inverse
problem solution in the form~(\ref{eq:invb}) this existence criterion
should be understood as follows. If for all $n$ for a given $\eta$
it is possible to find a finite $m=M_{n}$, starting from which the
condition~(\ref{eq:conv_criterion}) is satisfied, then the solution
to the inverse problem exists. Note that it follows from the obtained
criterion that for each $Q_{m}$ there exists some $\eta_{cr}$ below
which the solution in the form of the inverse Bernoulli transform
does not exist.

\section{Examples of application of the existence criterion}

\subsection{Poisson distribution}

\noindent Let $Q_{m}$ be the Poisson distribution:

\begin{equation}
Q_{m}=\frac{\left(\overline{m}\right)^{m}}{m!}e^{-\overline{m}},\label{eq:poisson}
\end{equation}
where $\overline{m}$ is the mean number of photocounts.

The stability criterion~(\ref{eq:conv_criterion}) for the distribution~(\ref{eq:poisson})
is written as:

\begin{equation}
\frac{\left(\overline{m}\right)^{m+1}}{\left(m+1\right)!}e^{-\overline{m}}<\left(1-\frac{n}{m+1}\right)\frac{\eta}{1-\eta}\frac{\left(\overline{m}\right)^{m}}{m!}e^{-\overline{m}}.\label{eq:10}
\end{equation}
From the relation~(\ref{eq:10}) it follows that the inequality is
satisfied if $m>M_{n}=n-1+\left(1-\eta\right)\eta^{-1}\overline{m}$.
Hence it follows that for any given $n$ and $\eta$ there exists
$M_{n}$, starting from which the existence criterion is fulfilled.
It means that in the case of the Poisson distribution the solution
obtained by inverse Bernoulli transform exists for arbitrary $\eta$.

\subsection{Compound Poisson distribution}

\begin{figure}[b]
\centering\includegraphics[width=1\linewidth]{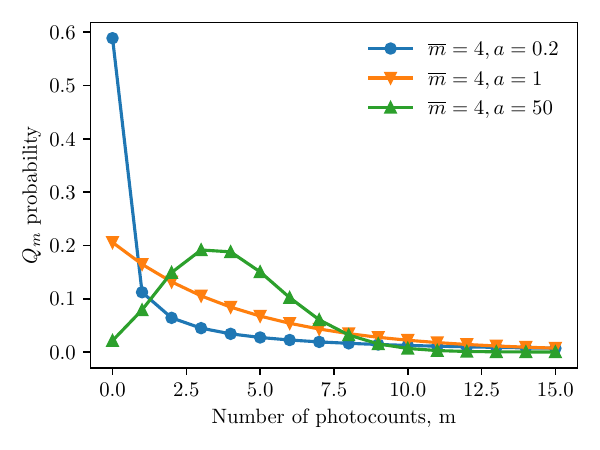}
\caption{\label{fig:2} Compound Poisson distributions with $\overline{m}=4$,
$a=0.2,1,50$.}
\end{figure}

Let $Q_{m}$ be the compound Poisson distribution:

\begin{equation}
Q_{m}=\frac{\Gamma(a+m)}{m!\Gamma(a)}\left(\frac{\overline{m}}{a}\right)^{m}\frac{1}{(1+\overline{m}/a)^{m+a}},\label{eq:comp_poisson}
\end{equation}
where $\overline{m}$ is the mean number of photocounts, $a$ is a
clusterization (or bunching) parameter. Using~(\ref{eq:comp_poisson})
we can describe a wide class of photocount distributions~\cite{bogdanov2016study}.
As shown from Fig.~\ref{fig:2}, this distribution strongly depends
on the $a$ value. If $a\rightarrow\infty$ it goes to the Poisson
distribution, if $a=1$ it coincides with the thermal one. Also, it
has a physical meaning if $0<a<1$ and for negative integers if $\overline{m}\leq-a$.
But for negative $a$ the distribution becomes finite and as shown
above all series~(\ref{eq:invb}) converge. So, problems of convergence
can arise only for $a>0$.

Writing down the existence criterion~(\ref{eq:conv_criterion}) for
the distribution~(\ref{eq:comp_poisson}) and taking into account
that $\Gamma(a+m+1)=\left(a+m\right)\Gamma(a+m)$, we arrive to inequality

\begin{equation}
m\left(\frac{\eta}{1-\eta}-\frac{\overline{m}}{a+\overline{m}}\right)>\frac{\eta\left(n-1\right)}{1-\eta}+\frac{\overline{m}a}{a+\overline{m}}.\label{eq:m_cond_comp_poisson}
\end{equation}

Further transformation of inequality~(\ref{eq:m_cond_comp_poisson})
depends on the sign of $\xi=\eta\left(1-\eta\right)^{-1}-\overline{m}\left(a+\overline{m}\right)^{-1}$.

If $\xi>0$, then the inequality~(\ref{eq:m_cond_comp_poisson})
can be written as 
\begin{align}
m>M_{n}=\left(\frac{\eta\left(n-1\right)}{1-\eta}+\frac{\overline{m}a}{a+\overline{m}}\right)\times\label{eq:m_cond_comp_poisson_corr}\\
\left(\frac{\eta}{1-\eta}-\frac{\overline{m}}{a+\overline{m}}\right)^{-1}.\nonumber 
\end{align}

Note that for $n=0$ the right-hand side of the inequality~(\ref{eq:m_cond_comp_poisson_corr})
for small $a$ can be negative, but it will hold for any $m$ and
we can put $M_{n}=0$.

If $\xi<0$, then the inequality~(\ref{eq:m_cond_comp_poisson})
leads to an upper constraint of $m$, which means that $M_{n}$ does
not exist.

Thus, the condition for the existence of $M_{n}$ is the fulfillment
of the condition $\xi>0$, from which it immediately follows that
\begin{equation}
\eta_{cr}=\left(\frac{a}{\overline{m}}+2\right)^{-1}.\label{eq:eta_cr}
\end{equation}

As is known, the compound Poisson distribution transforms into the
usual Poisson distribution for $a\rightarrow\infty$. With this $a$
the critical quantum efficiency $\eta_{cr}=0$, i.e. the solution
exists for any $\eta$, which coincides with the conclusion obtained
above when analyzing the usual Poisson distribution.

The expression~(\ref{eq:eta_cr}) also generalizes the special case
of the thermal distribution given in~\cite{kiss1995compensation}
as an example of the possibility of the existence of unstable solutions.
The compound Poisson distribution transforms into a thermal distribution
at $a=1$. Substituting this value into~(\ref{eq:eta_cr}) we get
$\eta_{cr}=\left(\overline{m}^{-1}+2\right)^{-1}$, which coincides
with the value obtained in~\cite{kiss1995compensation} from fundamentally
different considerations.

The obtained coincidences are an additional confirmation of the correctness
of the existence criterion obtained in this paper.

\section{Conclusions}

As was discussed above, non-ideal quantum efficiency $\eta<1$ leads
to a number of problems when trying to reconstruct photon-number distribution
$P_{n}$ from the measured photocount distribution $Q_{m}$. These
problems must be taken into account in practical implementation of
the recovery procedure. In the present paper we have shown that for
$\eta<0.5$ in the case of an infinite distribution of photons, the
problems associated with solving the inverse problem of photocount
statistics in the form~(\ref{eq:minverse}) are not in the instability
of the behavior of its solution, but in the fact that a solution in
the form of the inverse Bernoulli transform does not always exist.

Also we have found the criterion that allows us to determine whether
or not there exists a solution to the inverse problem of photocount
statistics in the form of the inverse Bernoulli transform for arbitrary
types of infinite photocount distributions for any $\eta\in(0,1]$.
According to the obtained criterion, it becomes possible for each
$Q_{m}$ to determine the minimum possible $\eta=\eta_{cr}$, below
which the reconstruction of the infinite $P_{n}$ distribution by
inverse Bernoulli transform becomes impossible.

The results obtained for finite distributions also seems to us rather
important since it can positively influence the development of numerical
methods for solving the inverse problem of photocount statistics.
The obtained results show that normalization of the experimentally
obtained distribution of photocounts is insufficient to obtain the
correct solution using the inverse Bernoulli transform. It is necessary
to additionally ensure that probabilities $Q_{m}$ lie inside the
$Q$-simplex. This condition indicates a new way of developing algorithms
for solving the inverse problem of photocount statistics.

\section{Declaration of competing interest}

The authors declare that they have no known competing financial interests
or personal relationships that could have appeared to influence the
work reported in this paper.

\section{Funding}

This work is supported by the Russian Science Foundation under grant
No. 21-12-00155.

\section{Acknowledgment}

The authors are grateful to Dr. Elena Khasina and Prof. Elena Suetnova
for fruitful discussions on mathematical issues.

\input{convergencePRA.bbl}

\end{document}

%% file: convergencePRA.bbl
%

%% file: convergencePRA.bbl
\begin{thebibliography}{36}%
\makeatletter
\providecommand \@ifxundefined [1]{%
 \@ifx{#1\undefined}
}%
\providecommand \@ifnum [1]{%
 \ifnum #1\expandafter \@firstoftwo
 \else \expandafter \@secondoftwo
 \fi
}%
\providecommand \@ifx [1]{%
 \ifx #1\expandafter \@firstoftwo
 \else \expandafter \@secondoftwo
 \fi
}%
\providecommand \natexlab [1]{#1}%
\providecommand \enquote  [1]{``#1''}%
\providecommand \bibnamefont  [1]{#1}%
\providecommand \bibfnamefont [1]{#1}%
\providecommand \citenamefont [1]{#1}%
\providecommand \href@noop [0]{\@secondoftwo}%
\providecommand \href [0]{\begingroup \@sanitize@url \@href}%
\providecommand \@href[1]{\@@startlink{#1}\@@href}%
\providecommand \@@href[1]{\endgroup#1\@@endlink}%
\providecommand \@sanitize@url [0]{\catcode `\\12\catcode `\$12\catcode
  `\&12\catcode `\#12\catcode `\^12\catcode `\_12\catcode `\%12\relax}%
\providecommand \@@startlink[1]{}%
\providecommand \@@endlink[0]{}%
\providecommand \url  [0]{\begingroup\@sanitize@url \@url }%
\providecommand \@url [1]{\endgroup\@href {#1}{\urlprefix }}%
\providecommand \urlprefix  [0]{URL }%
\providecommand \Eprint [0]{\href }%
\providecommand \doibase [0]{https://doi.org/}%
\providecommand \selectlanguage [0]{\@gobble}%
\providecommand \bibinfo  [0]{\@secondoftwo}%
\providecommand \bibfield  [0]{\@secondoftwo}%
\providecommand \translation [1]{[#1]}%
\providecommand \BibitemOpen [0]{}%
\providecommand \bibitemStop [0]{}%
\providecommand \bibitemNoStop [0]{.\EOS\space}%
\providecommand \EOS [0]{\spacefactor3000\relax}%
\providecommand \BibitemShut  [1]{\csname bibitem#1\endcsname}%
\let\auto@bib@innerbib\@empty
\bibitem [{\citenamefont {Britun}\ and\ \citenamefont
  {Nikiforov}(2018)}]{britun2018photon}%
  \BibitemOpen
  \bibfield  {author} {\bibinfo {author} {\bibfnamefont {N.}~\bibnamefont
  {Britun}}\ and\ \bibinfo {author} {\bibfnamefont {A.}~\bibnamefont
  {Nikiforov}},\ }\href@noop {} {\emph {\bibinfo {title} {Photon Counting:
  Fundamentals and Applications}}}\ (\bibinfo  {publisher} {BoD--Books on
  Demand},\ \bibinfo {year} {2018})\BibitemShut {NoStop}%
\bibitem [{\citenamefont {Eisaman}\ \emph {et~al.}(2011)\citenamefont
  {Eisaman}, \citenamefont {Fan}, \citenamefont {Migdall},\ and\ \citenamefont
  {Polyakov}}]{eisaman2011invited}%
  \BibitemOpen
  \bibfield  {author} {\bibinfo {author} {\bibfnamefont {M.~D.}\ \bibnamefont
  {Eisaman}}, \bibinfo {author} {\bibfnamefont {J.}~\bibnamefont {Fan}},
  \bibinfo {author} {\bibfnamefont {A.}~\bibnamefont {Migdall}},\ and\ \bibinfo
  {author} {\bibfnamefont {S.~V.}\ \bibnamefont {Polyakov}},\ }\bibfield
  {title} {\bibinfo {title} {Invited review article: Single-photon sources and
  detectors},\ }\href {https://doi.org/10.1063/1.3610677} {\bibfield  {journal}
  {\bibinfo  {journal} {Rev. Sci. Instr.}\ }\textbf {\bibinfo {volume} {82}},\
  \bibinfo {pages} {071101} (\bibinfo {year} {2011})}\BibitemShut {NoStop}%
\bibitem [{\citenamefont {Morel}\ and\ \citenamefont
  {Saha}(2005)}]{morel2005100}%
  \BibitemOpen
  \bibfield  {author} {\bibinfo {author} {\bibfnamefont {S.}~\bibnamefont
  {Morel}}\ and\ \bibinfo {author} {\bibfnamefont {S.~K.}\ \bibnamefont
  {Saha}},\ }\bibinfo {title} {100 years of photon-counting: The quest for the
  perfect eye},\ in\ \href@noop {} {\emph {\bibinfo {booktitle} {21st Century
  Astrophysics}}},\ \bibinfo {series and number} {21st Century Astrophysics}\
  (\bibinfo  {publisher} {Anita Publications, Delhi},\ \bibinfo {year} {2005})\
  pp.\ \bibinfo {pages} {237--257}\BibitemShut {NoStop}%
\bibitem [{\citenamefont {Wick}(1989)}]{wick1989photon}%
  \BibitemOpen
  \bibfield  {author} {\bibinfo {author} {\bibfnamefont {R.}~\bibnamefont
  {Wick}},\ }\bibfield  {title} {\bibinfo {title} {Photon counting imaging:
  applications in biomedical research},\ }\href@noop {} {\bibfield  {journal}
  {\bibinfo  {journal} {Biotechniques}\ }\textbf {\bibinfo {volume} {7}},\
  \bibinfo {pages} {262} (\bibinfo {year} {1989})}\BibitemShut {NoStop}%
\bibitem [{\citenamefont {Hadfield}(2009)}]{hadfield2009single}%
  \BibitemOpen
  \bibfield  {author} {\bibinfo {author} {\bibfnamefont {R.~H.}\ \bibnamefont
  {Hadfield}},\ }\bibfield  {title} {\bibinfo {title} {Single-photon detectors
  for optical quantum information applications},\ }\href
  {https://doi.org/10.1038/nphoton.2009.230} {\bibfield  {journal} {\bibinfo
  {journal} {Nat. Photonics}\ }\textbf {\bibinfo {volume} {3}},\ \bibinfo
  {pages} {696} (\bibinfo {year} {2009})}\BibitemShut {NoStop}%
\bibitem [{\citenamefont {Taguchi}\ and\ \citenamefont
  {Iwanczyk}(2013)}]{taguchi2013vision}%
  \BibitemOpen
  \bibfield  {author} {\bibinfo {author} {\bibfnamefont {K.}~\bibnamefont
  {Taguchi}}\ and\ \bibinfo {author} {\bibfnamefont {J.~S.}\ \bibnamefont
  {Iwanczyk}},\ }\bibfield  {title} {\bibinfo {title} {Vision 20/20: single
  photon counting x-ray detectors in medical imaging},\ }\bibfield  {journal}
  {\bibinfo  {journal} {Med. Phys.}\ }\textbf {\bibinfo {volume} {40}},\ \href
  {https://doi.org/10.1118/1.4820371} {10.1118/1.4820371} (\bibinfo {year}
  {2013})\BibitemShut {NoStop}%
\bibitem [{\citenamefont {Chunnilall}\ \emph {et~al.}(2014)\citenamefont
  {Chunnilall}, \citenamefont {Degiovanni}, \citenamefont {K{\"u}ck},
  \citenamefont {M{\"u}ller},\ and\ \citenamefont
  {Sinclair}}]{chunnilall2014metrology}%
  \BibitemOpen
  \bibfield  {author} {\bibinfo {author} {\bibfnamefont {C.~J.}\ \bibnamefont
  {Chunnilall}}, \bibinfo {author} {\bibfnamefont {I.~P.}\ \bibnamefont
  {Degiovanni}}, \bibinfo {author} {\bibfnamefont {S.}~\bibnamefont
  {K{\"u}ck}}, \bibinfo {author} {\bibfnamefont {I.}~\bibnamefont
  {M{\"u}ller}},\ and\ \bibinfo {author} {\bibfnamefont {A.~G.}\ \bibnamefont
  {Sinclair}},\ }\bibfield  {title} {\bibinfo {title} {Metrology of
  single-photon sources and detectors: a review},\ }\href
  {https://doi.org/10.1117/1.oe.53.8.081910} {\bibfield  {journal} {\bibinfo
  {journal} {Opt. Eng.}\ }\textbf {\bibinfo {volume} {53}},\ \bibinfo {pages}
  {081910} (\bibinfo {year} {2014})}\BibitemShut {NoStop}%
\bibitem [{\citenamefont {Flohr}\ \emph {et~al.}(2020)\citenamefont {Flohr},
  \citenamefont {Petersilka}, \citenamefont {Henning}, \citenamefont
  {Ulzheimer}, \citenamefont {Ferda},\ and\ \citenamefont
  {Schmidt}}]{flohr2020photon}%
  \BibitemOpen
  \bibfield  {author} {\bibinfo {author} {\bibfnamefont {T.}~\bibnamefont
  {Flohr}}, \bibinfo {author} {\bibfnamefont {M.}~\bibnamefont {Petersilka}},
  \bibinfo {author} {\bibfnamefont {A.}~\bibnamefont {Henning}}, \bibinfo
  {author} {\bibfnamefont {S.}~\bibnamefont {Ulzheimer}}, \bibinfo {author}
  {\bibfnamefont {J.}~\bibnamefont {Ferda}},\ and\ \bibinfo {author}
  {\bibfnamefont {B.}~\bibnamefont {Schmidt}},\ }\bibfield  {title} {\bibinfo
  {title} {Photon-counting ct review},\ }\href
  {https://doi.org/10.1016/j.ejmp.2020.10.030} {\bibfield  {journal} {\bibinfo
  {journal} {Phys. Medica}\ }\textbf {\bibinfo {volume} {79}},\ \bibinfo
  {pages} {126} (\bibinfo {year} {2020})}\BibitemShut {NoStop}%
\bibitem [{\citenamefont {Banaszek}\ and\ \citenamefont
  {W{\'o}dkiewicz}(1996)}]{banaszek1996direct}%
  \BibitemOpen
  \bibfield  {author} {\bibinfo {author} {\bibfnamefont {K.}~\bibnamefont
  {Banaszek}}\ and\ \bibinfo {author} {\bibfnamefont {K.}~\bibnamefont
  {W{\'o}dkiewicz}},\ }\bibfield  {title} {\bibinfo {title} {Direct probing of
  quantum phase space by photon counting},\ }\href
  {https://doi.org/10.1103/physrevlett.76.4344} {\bibfield  {journal} {\bibinfo
   {journal} {PRL}\ }\textbf {\bibinfo {volume} {76}},\ \bibinfo {pages} {4344}
  (\bibinfo {year} {1996})}\BibitemShut {NoStop}%
\bibitem [{\citenamefont {Banaszek}\ \emph {et~al.}(1999)\citenamefont
  {Banaszek}, \citenamefont {Radzewicz}, \citenamefont {W{\'o}dkiewicz},\ and\
  \citenamefont {Krasi{\'n}ski}}]{banaszek1999direct}%
  \BibitemOpen
  \bibfield  {author} {\bibinfo {author} {\bibfnamefont {K.}~\bibnamefont
  {Banaszek}}, \bibinfo {author} {\bibfnamefont {C.}~\bibnamefont {Radzewicz}},
  \bibinfo {author} {\bibfnamefont {K.}~\bibnamefont {W{\'o}dkiewicz}},\ and\
  \bibinfo {author} {\bibfnamefont {J.}~\bibnamefont {Krasi{\'n}ski}},\
  }\bibfield  {title} {\bibinfo {title} {Direct measurement of the wigner
  function by photon counting},\ }\href
  {https://doi.org/10.1103/physreva.60.674} {\bibfield  {journal} {\bibinfo
  {journal} {Phys. Rev. A}\ }\textbf {\bibinfo {volume} {60}},\ \bibinfo
  {pages} {674} (\bibinfo {year} {1999})},\ \Eprint
  {https://arxiv.org/abs/quant-ph/9903027} {quant-ph/9903027} \BibitemShut
  {NoStop}%
\bibitem [{\citenamefont {Nehra}\ \emph {et~al.}(2019)\citenamefont {Nehra},
  \citenamefont {Win}, \citenamefont {Eaton}, \citenamefont {Sridhar},
  \citenamefont {Shahrokhshahi}, \citenamefont {Gerrits}, \citenamefont {Lita},
  \citenamefont {Nam},\ and\ \citenamefont {Pfister}}]{nehra2019state}%
  \BibitemOpen
  \bibfield  {author} {\bibinfo {author} {\bibfnamefont {R.}~\bibnamefont
  {Nehra}}, \bibinfo {author} {\bibfnamefont {A.}~\bibnamefont {Win}}, \bibinfo
  {author} {\bibfnamefont {M.}~\bibnamefont {Eaton}}, \bibinfo {author}
  {\bibfnamefont {N.}~\bibnamefont {Sridhar}}, \bibinfo {author} {\bibfnamefont
  {R.}~\bibnamefont {Shahrokhshahi}}, \bibinfo {author} {\bibfnamefont
  {T.}~\bibnamefont {Gerrits}}, \bibinfo {author} {\bibfnamefont
  {A.}~\bibnamefont {Lita}}, \bibinfo {author} {\bibfnamefont {S.~W.}\
  \bibnamefont {Nam}},\ and\ \bibinfo {author} {\bibfnamefont {O.}~\bibnamefont
  {Pfister}},\ }\bibfield  {title} {\bibinfo {title} {State-independent quantum
  tomography of a single-photon state by photon-number-resolving measurements}\
  }\textbf {\bibinfo {volume} {6}},\ \href
  {https://doi.org/10.1364/OPTICA.6.001356} {10.1364/OPTICA.6.001356} (\bibinfo
  {year} {2019}),\ \Eprint {https://arxiv.org/abs/1906.02093} {1906.02093}
  \BibitemShut {NoStop}%
\bibitem [{\citenamefont {Ortolano}\ \emph {et~al.}(2021)\citenamefont
  {Ortolano}, \citenamefont {Losero}, \citenamefont {Pirandola}, \citenamefont
  {Genovese},\ and\ \citenamefont {Ruo-Berchera}}]{ortolano2021experimental}%
  \BibitemOpen
  \bibfield  {author} {\bibinfo {author} {\bibfnamefont {G.}~\bibnamefont
  {Ortolano}}, \bibinfo {author} {\bibfnamefont {E.}~\bibnamefont {Losero}},
  \bibinfo {author} {\bibfnamefont {S.}~\bibnamefont {Pirandola}}, \bibinfo
  {author} {\bibfnamefont {M.}~\bibnamefont {Genovese}},\ and\ \bibinfo
  {author} {\bibfnamefont {I.}~\bibnamefont {Ruo-Berchera}},\ }\bibfield
  {title} {\bibinfo {title} {Experimental quantum reading with photon
  counting},\ }\href {https://doi.org/10.1126/sciadv.abc7796} {\bibfield
  {journal} {\bibinfo  {journal} {Sci. Adv.}\ }\textbf {\bibinfo {volume}
  {7}},\ \bibinfo {pages} {eabc7796} (\bibinfo {year} {2021})}\BibitemShut
  {NoStop}%
\bibitem [{\citenamefont {Wolf}\ and\ \citenamefont
  {Mehta}(1964)}]{wolf1964determination}%
  \BibitemOpen
  \bibfield  {author} {\bibinfo {author} {\bibfnamefont {E.}~\bibnamefont
  {Wolf}}\ and\ \bibinfo {author} {\bibfnamefont {C.~L.}\ \bibnamefont
  {Mehta}},\ }\bibfield  {title} {\bibinfo {title} {Determination of the
  statistical properties of light from photoelectric measurements},\ }\href
  {https://doi.org/10.1103/physrevlett.13.705} {\bibfield  {journal} {\bibinfo
  {journal} {PRL}\ }\textbf {\bibinfo {volume} {13}},\ \bibinfo {pages} {705}
  (\bibinfo {year} {1964})}\BibitemShut {NoStop}%
\bibitem [{\citenamefont {B\'{e}dard}(1967)}]{bedard1967light}%
  \BibitemOpen
  \bibfield  {author} {\bibinfo {author} {\bibfnamefont {G.}~\bibnamefont
  {B\'{e}dard}},\ }\bibfield  {title} {\bibinfo {title} {Analysis of light
  fluctuations from photon counting statistics},\ }\href
  {https://doi.org/10.1364/JOSA.57.001201} {\bibfield  {journal} {\bibinfo
  {journal} {J. Opt. Soc. Am.}\ }\textbf {\bibinfo {volume} {57}},\ \bibinfo
  {pages} {1201} (\bibinfo {year} {1967})}\BibitemShut {NoStop}%
\bibitem [{\citenamefont {Sultani}\ \emph {et~al.}(1995)\citenamefont
  {Sultani}, \citenamefont {Aime},\ and\ \citenamefont
  {Lant{\'e}ri}}]{sultani1995inverse}%
  \BibitemOpen
  \bibfield  {author} {\bibinfo {author} {\bibfnamefont {F.}~\bibnamefont
  {Sultani}}, \bibinfo {author} {\bibfnamefont {C.}~\bibnamefont {Aime}},\ and\
  \bibinfo {author} {\bibfnamefont {H.}~\bibnamefont {Lant{\'e}ri}},\
  }\bibfield  {title} {\bibinfo {title} {Inverse poisson transform using
  pad{\'e} approximants. applications to speckle interferometry in astronomy},\
  }\href {https://doi.org/10.1088/0963-9659/4/2/005} {\bibfield  {journal}
  {\bibinfo  {journal} {Pure Appl. Opt.}\ }\textbf {\bibinfo {volume} {4}},\
  \bibinfo {pages} {89} (\bibinfo {year} {1995})}\BibitemShut {NoStop}%
\bibitem [{\citenamefont {Earnshaw}\ and\ \citenamefont
  {Haughey}(1996)}]{earnshaw1996inversion}%
  \BibitemOpen
  \bibfield  {author} {\bibinfo {author} {\bibfnamefont {J.}~\bibnamefont
  {Earnshaw}}\ and\ \bibinfo {author} {\bibfnamefont {D.}~\bibnamefont
  {Haughey}},\ }\bibfield  {title} {\bibinfo {title} {Inversion of the poisson
  transform using proportionally spaced cubic b-splines},\ }\href
  {https://doi.org/10.1063/1.1147540} {\bibfield  {journal} {\bibinfo
  {journal} {Rev. Sci. Instr.}\ }\textbf {\bibinfo {volume} {67}},\ \bibinfo
  {pages} {4387} (\bibinfo {year} {1996})}\BibitemShut {NoStop}%
\bibitem [{\citenamefont {Mandel}(1958)}]{mandel1958fluctuations}%
  \BibitemOpen
  \bibfield  {author} {\bibinfo {author} {\bibfnamefont {L.}~\bibnamefont
  {Mandel}},\ }\bibfield  {title} {\bibinfo {title} {Fluctuations of photon
  beams and their correlations},\ }\href
  {https://doi.org/10.1088/0370-1328/72/6/312} {\bibfield  {journal} {\bibinfo
  {journal} {Proc. Phys. Soc.}\ }\textbf {\bibinfo {volume} {72}},\ \bibinfo
  {pages} {1037} (\bibinfo {year} {1958})}\BibitemShut {NoStop}%
\bibitem [{\citenamefont {Akhmanov}\ \emph {et~al.}(1981)\citenamefont
  {Akhmanov}, \citenamefont {D’yakov},\ and\ \citenamefont
  {Chirkin}}]{akhmanov1981introduction}%
  \BibitemOpen
  \bibfield  {author} {\bibinfo {author} {\bibfnamefont {S.~A.}\ \bibnamefont
  {Akhmanov}}, \bibinfo {author} {\bibfnamefont {Y.~E.}\ \bibnamefont
  {D’yakov}},\ and\ \bibinfo {author} {\bibfnamefont {A.~S.}\ \bibnamefont
  {Chirkin}},\ }\href@noop {} {\emph {\bibinfo {title} {Introduction to
  statistical radiophysics and optics}}}\ (\bibinfo  {publisher} {Nauka,
  Moscow},\ \bibinfo {year} {1981})\BibitemShut {NoStop}%
\bibitem [{\citenamefont {Vogel}\ and\ \citenamefont
  {Welsch}(2006)}]{vogel2006quantum}%
  \BibitemOpen
  \bibfield  {author} {\bibinfo {author} {\bibfnamefont {W.}~\bibnamefont
  {Vogel}}\ and\ \bibinfo {author} {\bibfnamefont {D.-G.}\ \bibnamefont
  {Welsch}},\ }\href {https://doi.org/10.1002/3527608524} {\emph {\bibinfo
  {title} {Quantum Optics}}}\ (\bibinfo  {publisher} {Wiley},\ \bibinfo {year}
  {2006})\BibitemShut {NoStop}%
\bibitem [{\citenamefont {Lee}(1993)}]{lee1993external}%
  \BibitemOpen
  \bibfield  {author} {\bibinfo {author} {\bibfnamefont {C.~T.}\ \bibnamefont
  {Lee}},\ }\bibfield  {title} {\bibinfo {title} {External photodetection of
  cavity radiation},\ }\href {https://doi.org/10.1103/physreva.48.2285}
  {\bibfield  {journal} {\bibinfo  {journal} {Phys. Rev. A}\ }\textbf {\bibinfo
  {volume} {48}},\ \bibinfo {pages} {2285} (\bibinfo {year}
  {1993})}\BibitemShut {NoStop}%
\bibitem [{\citenamefont {Kiss}\ \emph {et~al.}(1995)\citenamefont {Kiss},
  \citenamefont {Herzog},\ and\ \citenamefont
  {Leonhardt}}]{kiss1995compensation}%
  \BibitemOpen
  \bibfield  {author} {\bibinfo {author} {\bibfnamefont {T.}~\bibnamefont
  {Kiss}}, \bibinfo {author} {\bibfnamefont {U.}~\bibnamefont {Herzog}},\ and\
  \bibinfo {author} {\bibfnamefont {U.}~\bibnamefont {Leonhardt}},\ }\bibfield
  {title} {\bibinfo {title} {Compensation of losses in photodetection and in
  quantum-state measurements},\ }\href
  {https://doi.org/10.1103/physreva.52.2433} {\bibfield  {journal} {\bibinfo
  {journal} {Phys. Rev. A}\ }\textbf {\bibinfo {volume} {52}},\ \bibinfo
  {pages} {2433} (\bibinfo {year} {1995})}\BibitemShut {NoStop}%
\bibitem [{\citenamefont {Herzog}(1996{\natexlab{a}})}]{herzog1996loss}%
  \BibitemOpen
  \bibfield  {author} {\bibinfo {author} {\bibfnamefont {U.}~\bibnamefont
  {Herzog}},\ }\bibfield  {title} {\bibinfo {title} {Loss-error compensation in
  quantum-state measurements and the solution of the time-reversed damping
  equation},\ }\href {https://doi.org/10.1103/physreva.53.1245} {\bibfield
  {journal} {\bibinfo  {journal} {Phys. Rev. A}\ }\textbf {\bibinfo {volume}
  {53}},\ \bibinfo {pages} {1245} (\bibinfo {year}
  {1996}{\natexlab{a}})}\BibitemShut {NoStop}%
\bibitem [{\citenamefont {Messin}\ \emph {et~al.}(2009)\citenamefont {Messin},
  \citenamefont {Sanders}, \citenamefont {Petrosyan},\ and\ \citenamefont
  {Rarity}}]{messin2009few}%
  \BibitemOpen
  \bibfield  {author} {\bibinfo {author} {\bibfnamefont {G.}~\bibnamefont
  {Messin}}, \bibinfo {author} {\bibfnamefont {B.~C.}\ \bibnamefont {Sanders}},
  \bibinfo {author} {\bibfnamefont {D.}~\bibnamefont {Petrosyan}},\ and\
  \bibinfo {author} {\bibfnamefont {J.}~\bibnamefont {Rarity}},\ }\bibfield
  {title} {\bibinfo {title} {Few-photon optics},\ }\href
  {https://doi.org/10.1088/0953-4075/42/11/110201} {\bibfield  {journal}
  {\bibinfo  {journal} {J. Phys. B: At. Mol. Opt. Phys.}\ }\textbf {\bibinfo
  {volume} {42}},\ \bibinfo {pages} {110201} (\bibinfo {year}
  {2009})}\BibitemShut {NoStop}%
\bibitem [{\citenamefont {Pathak}\ and\ \citenamefont
  {Ghatak}(2018)}]{pathak2018classical}%
  \BibitemOpen
  \bibfield  {author} {\bibinfo {author} {\bibfnamefont {A.}~\bibnamefont
  {Pathak}}\ and\ \bibinfo {author} {\bibfnamefont {A.}~\bibnamefont
  {Ghatak}},\ }\bibfield  {title} {\bibinfo {title} {Classical light vs.
  nonclassical light: characterizations and interesting applications},\
  }\href@noop {} {\bibfield  {journal} {\bibinfo  {journal} {J. Electromagnet.
  Wave.}\ }\textbf {\bibinfo {volume} {32}},\ \bibinfo {pages} {229} (\bibinfo
  {year} {2018})},\ \Eprint {https://arxiv.org/abs/1705.00650} {1705.00650}
  \BibitemShut {NoStop}%
\bibitem [{\citenamefont {Gundacker}\ and\ \citenamefont
  {Heering}(2020)}]{gundacker2020silicon}%
  \BibitemOpen
  \bibfield  {author} {\bibinfo {author} {\bibfnamefont {S.}~\bibnamefont
  {Gundacker}}\ and\ \bibinfo {author} {\bibfnamefont {A.}~\bibnamefont
  {Heering}},\ }\bibfield  {title} {\bibinfo {title} {The silicon
  photomultiplier: fundamentals and applications of a modern solid-state photon
  detector},\ }\href {https://doi.org/10.1088/1361-6560/ab7b2d} {\bibfield
  {journal} {\bibinfo  {journal} {Phys. Med. Biol.}\ }\textbf {\bibinfo
  {volume} {65}},\ \bibinfo {pages} {17TR01} (\bibinfo {year}
  {2020})}\BibitemShut {NoStop}%
\bibitem [{\citenamefont {Aitchison}\ and\ \citenamefont
  {Silvey}(1958)}]{aitchison1958maximum}%
  \BibitemOpen
  \bibfield  {author} {\bibinfo {author} {\bibfnamefont {J.}~\bibnamefont
  {Aitchison}}\ and\ \bibinfo {author} {\bibfnamefont {S.}~\bibnamefont
  {Silvey}},\ }\bibfield  {title} {\bibinfo {title} {Maximum-likelihood
  estimation of parameters subject to restraints},\ }\href@noop {} {\bibfield
  {journal} {\bibinfo  {journal} {The annals of mathematical Statistics}\
  }\textbf {\bibinfo {volume} {29}},\ \bibinfo {pages} {813} (\bibinfo {year}
  {1958})}\BibitemShut {NoStop}%
\bibitem [{\citenamefont {Zambra}\ and\ \citenamefont
  {Paris}(2006)}]{zambra2006reconstruction}%
  \BibitemOpen
  \bibfield  {author} {\bibinfo {author} {\bibfnamefont {G.}~\bibnamefont
  {Zambra}}\ and\ \bibinfo {author} {\bibfnamefont {M.~G.~A.}\ \bibnamefont
  {Paris}},\ }\bibfield  {title} {\bibinfo {title} {Reconstruction of
  photon-number distribution using low-performance photon counters},\
  }\href@noop {} {\bibfield  {journal} {\bibinfo  {journal} {Phys. Rev. A}\
  }\textbf {\bibinfo {volume} {74}},\ \bibinfo {pages} {063830} (\bibinfo
  {year} {2006})}\BibitemShut {NoStop}%
\bibitem [{\citenamefont {Zambra}\ \emph {et~al.}(2007)\citenamefont {Zambra},
  \citenamefont {Allevi}, \citenamefont {Andreoni}, \citenamefont {Bondani},\
  and\ \citenamefont {Paris}}]{zambra2007nontrivial}%
  \BibitemOpen
  \bibfield  {author} {\bibinfo {author} {\bibfnamefont {G.}~\bibnamefont
  {Zambra}}, \bibinfo {author} {\bibfnamefont {A.}~\bibnamefont {Allevi}},
  \bibinfo {author} {\bibfnamefont {A.}~\bibnamefont {Andreoni}}, \bibinfo
  {author} {\bibfnamefont {M.}~\bibnamefont {Bondani}},\ and\ \bibinfo {author}
  {\bibfnamefont {M.~G.~A.}\ \bibnamefont {Paris}},\ }\bibfield  {title}
  {\bibinfo {title} {Nontrivial photon statistics with low resolution-threshold
  photon counters},\ }\href@noop {} {\bibfield  {journal} {\bibinfo  {journal}
  {Int. J. of Quant. Info.}\ }\textbf {\bibinfo {volume} {5}},\ \bibinfo
  {pages} {305} (\bibinfo {year} {2007})}\BibitemShut {NoStop}%
\bibitem [{\citenamefont {Hlou{\v{s}}ek}\ \emph {et~al.}(2019)\citenamefont
  {Hlou{\v{s}}ek}, \citenamefont {Dudka}, \citenamefont {Straka},\ and\
  \citenamefont {Je{\v{z}}ek}}]{hlouvsek2019accurate}%
  \BibitemOpen
  \bibfield  {author} {\bibinfo {author} {\bibfnamefont {J.}~\bibnamefont
  {Hlou{\v{s}}ek}}, \bibinfo {author} {\bibfnamefont {M.}~\bibnamefont
  {Dudka}}, \bibinfo {author} {\bibfnamefont {I.}~\bibnamefont {Straka}},\ and\
  \bibinfo {author} {\bibfnamefont {M.}~\bibnamefont {Je{\v{z}}ek}},\
  }\bibfield  {title} {\bibinfo {title} {Accurate detection of arbitrary photon
  statistics},\ }\href@noop {} {\bibfield  {journal} {\bibinfo  {journal}
  {Physical review letters}\ }\textbf {\bibinfo {volume} {123}},\ \bibinfo
  {pages} {153604} (\bibinfo {year} {2019})}\BibitemShut {NoStop}%
\bibitem [{\citenamefont {Wunsche}(1990)}]{wunsche1990reconstruction}%
  \BibitemOpen
  \bibfield  {author} {\bibinfo {author} {\bibfnamefont {A.}~\bibnamefont
  {Wunsche}},\ }\bibfield  {title} {\bibinfo {title} {Reconstruction of
  operators from their normally ordered moments for a single boson mode},\
  }\href {https://doi.org/10.1088/0954-8998/2/6/004} {\bibfield  {journal}
  {\bibinfo  {journal} {Quantum Opt.}\ }\textbf {\bibinfo {volume} {2}},\
  \bibinfo {pages} {453} (\bibinfo {year} {1990})}\BibitemShut {NoStop}%
\bibitem [{\citenamefont {Herzog}(1996{\natexlab{b}})}]{herzog1996generating}%
  \BibitemOpen
  \bibfield  {author} {\bibinfo {author} {\bibfnamefont {U.}~\bibnamefont
  {Herzog}},\ }\bibfield  {title} {\bibinfo {title} {Generating-function
  approach to the moment problem for the density matrix of a single mode},\
  }\href {https://doi.org/10.1103/physreva.53.2889} {\bibfield  {journal}
  {\bibinfo  {journal} {Phys. Rev. A}\ }\textbf {\bibinfo {volume} {53}},\
  \bibinfo {pages} {2889} (\bibinfo {year} {1996}{\natexlab{b}})}\BibitemShut
  {NoStop}%
\bibitem [{\citenamefont {D’Ariano}\ and\ \citenamefont
  {Macchiavello}(1998)}]{d1998loss}%
  \BibitemOpen
  \bibfield  {author} {\bibinfo {author} {\bibfnamefont {G.}~\bibnamefont
  {D’Ariano}}\ and\ \bibinfo {author} {\bibfnamefont {C.}~\bibnamefont
  {Macchiavello}},\ }\bibfield  {title} {\bibinfo {title} {Loss-error
  compensation in quantum-state measurements},\ }\href@noop {} {\bibfield
  {journal} {\bibinfo  {journal} {Physical Review A}\ }\textbf {\bibinfo
  {volume} {57}},\ \bibinfo {pages} {3131} (\bibinfo {year}
  {1998})}\BibitemShut {NoStop}%
\bibitem [{\citenamefont {Kiss}\ \emph
  {et~al.}(1998{\natexlab{a}})\citenamefont {Kiss}, \citenamefont {Herzog},\
  and\ \citenamefont {Leonhardt}}]{kiss1998reply}%
  \BibitemOpen
  \bibfield  {author} {\bibinfo {author} {\bibfnamefont {T.}~\bibnamefont
  {Kiss}}, \bibinfo {author} {\bibfnamefont {U.}~\bibnamefont {Herzog}},\ and\
  \bibinfo {author} {\bibfnamefont {U.}~\bibnamefont {Leonhardt}},\ }\bibfield
  {title} {\bibinfo {title} {Reply to “loss-error compensation in
  quantum-state measurements”},\ }\href@noop {} {\bibfield  {journal}
  {\bibinfo  {journal} {Physical Review A}\ }\textbf {\bibinfo {volume} {57}},\
  \bibinfo {pages} {3134} (\bibinfo {year} {1998}{\natexlab{a}})}\BibitemShut
  {NoStop}%
\bibitem [{\citenamefont {D'Ariano}\ and\ \citenamefont
  {Macchiavello}(1997)}]{d1997comment}%
  \BibitemOpen
  \bibfield  {author} {\bibinfo {author} {\bibfnamefont {G.}~\bibnamefont
  {D'Ariano}}\ and\ \bibinfo {author} {\bibfnamefont {C.}~\bibnamefont
  {Macchiavello}},\ }\bibfield  {title} {\bibinfo {title} {Comment on"
  loss-error compensation in quantum-state measurements"},\ }\href@noop {}
  {\bibfield  {journal} {\bibinfo  {journal} {arXiv preprint quant-ph/9701009}\
  } (\bibinfo {year} {1997})}\BibitemShut {NoStop}%
\bibitem [{\citenamefont {Kiss}\ \emph
  {et~al.}(1998{\natexlab{b}})\citenamefont {Kiss}, \citenamefont {Herzog},\
  and\ \citenamefont {Leonhardt}}]{kiss1998replyoncomm}%
  \BibitemOpen
  \bibfield  {author} {\bibinfo {author} {\bibfnamefont {T.}~\bibnamefont
  {Kiss}}, \bibinfo {author} {\bibfnamefont {U.}~\bibnamefont {Herzog}},\ and\
  \bibinfo {author} {\bibfnamefont {U.}~\bibnamefont {Leonhardt}},\ }\bibfield
  {title} {\bibinfo {title} {Reply on the``comment on loss-error compensation
  in quantum-state measurements'''},\ }\href@noop {} {\bibfield  {journal}
  {\bibinfo  {journal} {arXiv preprint quant-ph/9801028}\ } (\bibinfo {year}
  {1998}{\natexlab{b}})}\BibitemShut {NoStop}%
\bibitem [{\citenamefont {Bogdanov}\ \emph {et~al.}(2016)\citenamefont
  {Bogdanov}, \citenamefont {Bogdanova}, \citenamefont {Katamadze},
  \citenamefont {Avosopyants},\ and\ \citenamefont
  {Lukichev}}]{bogdanov2016study}%
  \BibitemOpen
  \bibfield  {author} {\bibinfo {author} {\bibfnamefont {Y.~I.}\ \bibnamefont
  {Bogdanov}}, \bibinfo {author} {\bibfnamefont {N.}~\bibnamefont {Bogdanova}},
  \bibinfo {author} {\bibfnamefont {K.}~\bibnamefont {Katamadze}}, \bibinfo
  {author} {\bibfnamefont {G.}~\bibnamefont {Avosopyants}},\ and\ \bibinfo
  {author} {\bibfnamefont {V.}~\bibnamefont {Lukichev}},\ }\bibfield  {title}
  {\bibinfo {title} {Study of photon statistics using a compound poisson
  distribution and quadrature measurements},\ }\href
  {https://doi.org/10.3103/s8756699016050095} {\bibfield  {journal} {\bibinfo
  {journal} {Optoelectron. Instrument. Proc.}\ }\textbf {\bibinfo {volume}
  {52}},\ \bibinfo {pages} {475} (\bibinfo {year} {2016})}\BibitemShut
  {NoStop}%
\end{thebibliography}
